\documentstyle[aps,prd,eqsecnum,epsfig]{revtex}
\begin{document}

\draft
\title{\bf Inhomogeneities in Newtonian Cosmology \\and its Backreaction
to the
Evolution of the Universe }

\author{Takayuki Tatekawa
\thanks{e-mail:
tatekawa@gravity.phys.waseda.ac.jp}$^1$,
Momoko Suda
\thanks{e-mail:
suda@gravity.phys.waseda.ac.jp}$^2$,
Kei-ichi Maeda
\thanks{e-mail:
maeda@gravity.phys.waseda.ac.jp}$^{1,3,4}$,
and Hiroto Kubotani
\thanks{e-mail:
kubotani@cpsun3.b6.kanagawa-u.ac.jp}$^5$}
\address{$^1$ Department of Physics, Waseda
University, 3-4-1 Okubo, Shinjuku-ku, Tokyo 169-8555,
Japan}
\address{$^2$ Mainichi Shimbun, Hitotsubashi, Chiyoda-ku,
Tokyo, Japan}
\address{$^3$ Advanced Research Institute for Science and
Engineering,
Waseda University, Shinjuku, Tokyo 169-8555, Japan}
\address{$^4$ Advanced Institute for Complex Systems,
Waseda University, Shinjuku, Tokyo 169-8555, Japan}
\address{$^5$ Faculty of Engineering,
Kanagawa University, Yokohama 221-8686, Japan}
\date{\today}
\maketitle



\begin{abstract}
We study an effect of inhomogeneity of density distribution of the
Universe.
We propose a new Lagrangian perturbation theory with a backreaction
effect by inhomogeneity.
   The inhomogeneity affects
the expansion rate in a local domain and its own growing rate.
   We  numerically analyze a one-dimensional  plane-symmetric
model, and
calculate  the probability distribution functions (PDFs) of several
observed variables to discuss those statistical properties.
We find that the PDF of
pairwise peculiar velocity shows an effective difference
from the conventional Lagrangian approach, i.e. even in one-dimensional
plane symmetric case, the PDF approaches  an
exponential form in a small relative-velocity region, which agree with
the N-body simulation.
\end{abstract}


\section{Introduction}
The present Universe shows a variety of structures. How such
a structure is formed in the evolution of the Universe. One
of the most plausible explanations is that the nonlinear
dynamics of a self-gravitating system provides such a scale-free
structure during the evolution of the Universe.
In order to clarify whether such a dynamics really
gives an appropriate observed feature, we have so far  three approaches:
$N$-body simulation, the Eulerian perturbation approach and
the Lagrangian one.
Although the final answer for a structure formation would be
obtained by the $N$-body simulation, it may  be difficult to
obtain enough resolution to discuss a fine structure of the Universe such
as a scaling property. As for a perturbation approach, however, it is
just an approximation and will break down in a nonlinear regime, although
the Lagrangian approach would be better if we are interested in density
perturbations. This is just because
a density fluctuation $\delta$ and a peculiar velocity ${\bf
v}$ are perturbed
quantities in the Eulerian approach\cite{Peebles80}, while
a displacement of particles from uniform distribution is assumed to be
small in the Lagrangian
approach\cite{Zeldovich70,Buchert89,Buchert92,Catelan95,SahCol95}. Its
first order solution is the so-called Zel'dovich approximation (ZA)
\cite{Zeldovich70}. The Lagrangian approach is confirmed to be better than
the Eulerian approach by comparison of exact solutions in several
cases\cite{MunSahSta94,SahSha96,YosMatMor98}.
Therefore, we will adopt the Lagrangian approximation in this paper and
discuss about how to improve it.

In the standard approach of Newtonian cosmology, the global cosmological
parameters such as Hubble expansion rate and mean density are given first
by a solution of the Einstein equations, i.e. the
so-called Friedmann-Robertson-Walker (FRW) universe, which is an
isotropic and homogeneous spacetime.  According to
observation, however, a local structure in the Universe is definitely  not
homogeneous and isotropic\cite{GelHuc89}.
In the standard approach, the density
   averaged over the whole space (or a horizon scale)
is assumed to be the energy density of the FRW spacetime.
However, here the problems of
how to average inhomogeneous matter fluid and how to define an
averaged isotropic and homogeneous spacetime are arisen.
This problem has been discussed in the
framework of Newtonian cosmology and general relativity by many authors
\cite{BilFut91,Futamase96,RusSofKasBor97,StoHelTor99,Buchert00,BucEhl97,EhlBuc97}.
The most important point to discuss is how inhomogeneities affect the
expansion law of the Universe and whether such a backreaction affects
the evolution of the density perturbations.
This is our subject to discuss in this paper.
Although many authors proposed different
methods of averaging procedure in general relativity
\cite{BilFut91,Futamase96,RusSofKasBor97,StoHelTor99,Buchert00},
a serious problem  remains.
Different gauge choices make the problem more complicated.

Since a relativistic effect may not be so important,
in this paper, in order to avoid such a difficulty,
we discuss only an averaging procedure
in the Newtonian framework. Proposing a averaging procedure, which  is
defined by spatial average of physical quantities,  Buchert and
Ehlers  lead the averaged
Raychaudhuri's equation\cite{BucEhl97}. The equation describes how the
averaged expansion rate of domain with a finite volume evolves.
This equation has an additional term, which we
call a 'backreaction term' of  inhomogeneities on
averaged expansion.
Then, Buchert, Kerscher, and Sicka estimated the
backreaction term using the conventional Lagrangian perturbation
approach\cite{BucKerSic00}.  They first consider density perturbations in
Einstein-de Sitter (E-dS) Universe, then calculated the backreaction term
and solved the averaged Raychaudhuri's equation. They showed difference
between a cosmological parameter such as Hubble expansion rate in their
averaged model and that of the E-dS model. Although they included a
backreaction term to estimate the averaged variables in a local domain,
they used the perturbed quantities from   the E-dS Universe. In other
words, they have not take into account a backreaction on the evolution of
density perturbations.

Here we improve their approach, i.e. we include a backreaction effect
to averaged expansion rate and solve the averaged
Raychaudhuri's equation (the generalized Friedmann's equation) with
evolution equation of perturbations in the averaged domain.
In an averaged domain, the averaged density is either higher or lower than
that of the E-dS universe. This difference will change  the evolution of
density perturbations. In fact, if
the domain is overdense, growth rate of perturbations behaves
   as that in the closed universe, While, if it is
   underdense, it is just  like a
solution in the  open universe.

This paper is organized as follows:
In $\S 2$, we shortly derive the  generalized Friedmann
equation, following Ehlers and Buchert, and estimate a backreaction
term. We present our formalism in
$\S 3$. We show some relation between two Lagrangian
descriptions (the conventional one and ours) and give
our initial setting in \S. 4.
In \S 5, we
analyze a simple example,i.e. a plane-symmetric one-dimensional model, to
show the probability distribution of the observed quantities such as
Hubble parameter, or density
fluctuation, or peculiar velocity. The
conclusion and discussion follow in $\S 6$.

\section{Averaging of inhomogeneity}
\subsection{The generalized Friedmann equation}
In the Newtonian cosmology, the expansion of a domain is
influenced by  inhomogeneity inside the domain.
Such an effect may be evaluated by spatial integration of field variables
in  the Lagrangian domain, which evolves with matter fluid.
Hence in this paper we study fields averaged over a simply-connected
spatial Lagrangian domain ${\cal D}$ at time $t$, which evolved out of the
initial domain ${\cal D}_i$ at time $t_i$.
   The locally averaged scale factor
$a_{\cal D}$, depending on the content, shape and location of the
   domain ${\cal D}$, is defined by the volume of  domain
$V_{\cal D}(t) = |{\cal D}|$ and its initial volume $V_{{\cal D}_i}=|{\cal
   D}_i|
$ as
\begin{equation}
a_{\cal D} (t) \equiv \left ( \frac{V_{\cal D}(t)}{V_{{\cal D}_i}} \right
)^{1/3} \,.
\end{equation}

We define a spatial averaging  for any rank
   tensor field ${\cal A} = \{ A_{ij\cdot} ({\bf r}, t) \}$ by the
    volume integral normalized by the volume of the domain as
\begin{equation}
\left <{\cal A} \right >_{\cal D} (t)\equiv \frac{1}{V_{\cal D}(t)}
\int_{\cal D} {\rm d}^3 r \; {\cal A} ({\bf r}, t) \,.
\end{equation}
The average
density is then given by
\begin{equation}
\left < \rho \right >_{\cal D}\equiv  \frac{M_{\cal D}} {V_{\cal D}(t)} =
\frac{M_{\cal D}} {a_{\cal D}^3 V_{{\cal D}_i}} = \frac{1}{a_{\cal D}^3}
   \left <
\rho
\right >_{{\cal D}_i}(t_i)
\propto {a_{\cal D}}^{-3}\,,
\end{equation}
where  the total mass $M_{\cal D} =\int_{\cal D} {\rm d}^3 r \; {\cal
\rho} ({\bf r}, t) $ for a domain ${\cal D}$ is conserved.
Using this averaging, we can derive the generalized Friedmann
equation\cite{BucEhl97}, which we will shortly derive here.

We start  from the hydrodynamic
equations for a self-gravitating perfect fluid;
\begin{eqnarray}
\frac{\partial \rho}{\partial t}  + \nabla_{\bf r} \cdot
    (\rho {\bf u}) &=& 0 \label{eqn:conti-eq} \;, \\
\frac{\partial {\bf u}}{\partial t}  + ({\bf u}
\cdot \nabla_{\bf r}) {\bf u} &=& - \frac{1}{\rho} \nabla_{
\bf r} P  -
\nabla_{\bf r} \Phi \label{eqn:euler-eq} \;, \\
\nabla_{\bf r} \times {\bf g} &=& {\bf 0}
\label{eqn:poisson-eq1} \;, \\
\nabla_{\bf r} \cdot {\bf g}  &=& \Lambda- 4 \pi G \rho
\label{eqn:poisson-eq2} \;.
\end{eqnarray}
Although it is easy to extend the case with pressure of fluid, we
   consider just dust matter and ignore a pressure term.

To discuss inhomogeneity, a spatial derivative of field will play a
crucial role.
In particular, it turns out that the spatial derivative of velocity {\bf
u},  which is divided into three variables as
\begin{eqnarray}
u_{i,j} &=& \sigma_{ij} + \frac{1}{3} \delta_{ij} \theta +
\omega_{ij} \;, \nonumber \\
   &&{\rm with }  ~~\omega_{ij} = - \varepsilon_{ijk} \omega^k \;,
\end{eqnarray}
is very important, where
\begin{eqnarray}
\theta &=&
\nabla
\cdot {\bf u}~~~~~~~~~~~~~~~~~~~~~~~~~~~({\rm expansion}) \nonumber \\
\sigma_{ij} &=&
\frac{1}{2}(u_{i,j}+u_{j,i})-\frac{1}{3}\delta_{ij}\theta ~~~~~~~~({\rm
shear})
\nonumber \\
\mbox{\boldmath{$\omega$}} &=&
\frac{1}{2}
\nabla
\times {\bf u} ~~~~~~~~~~~~~~~~~~~~~~~~({\rm  rotation
}).
\end{eqnarray}
The magnitudes of shear $\sigma$ and of rotation $\omega$ are defined
by
\begin{equation}
\sigma=|\underline{\sigma}|\equiv\left(\frac{1}{2}\sigma_{ij}
\sigma_{ij}\right)^{1\over 2}\,,\qquad
\omega=|\mbox{\boldmath{$\omega$}}|=\left(\frac{1}{2}
\omega_{ij}\omega_{ij}\right)^{1\over 2} \;.
\end{equation}
Using these variables, we find the basic equations for density, shear and
rotation from Eqs. (\ref{eqn:conti-eq}), (\ref{eqn:euler-eq}) as
\begin{eqnarray}
\dot\rho&=&-\theta\rho \;, \label{eqn:rho}\\
\dot{\mbox{\boldmath{$\omega$}}}&=&-\frac{2}{3}\theta
\mbox{\boldmath{$\omega$}}+\underline{\sigma}
\cdot
\mbox{\boldmath{$\omega$}} \;, \label{eqn:omega}\\
\dot{\theta}&=&\Lambda-4\pi G \rho-\frac{1}{3} \theta^2+2(
\omega^2-\sigma^2) \;.
\label{eqn:theta}
\end{eqnarray}

The volume of a spatial domain ${\cal
D}$ described by its initial domain ${\cal D}_i$
through a transformation from the Eulerian coordinates ${\bf r}$ to the
Lagrangian one ${\bf q}$ as
\begin{equation}
V_{\cal D} = \int_{\cal D} {\rm d}^3 r =\int_{{\cal D}_i} {
\rm d}^3 q  J_r\; ,
\end{equation}
where $J_r$ is the Jacobian of the transformation
\begin{equation}
J_r = \det \left ( \frac{\partial r_i}{\partial q_j} \right )
\;.
\end{equation}
The change rate   of the volume
of a domain is then given as
\begin{eqnarray}
\frac{1}{V_{\cal D}} \frac{{\rm d} V_{\cal D}}{{\rm d} t} &=& \frac{1}{V_{\cal
D}}
\frac{\rm d}{{\rm d}t} \int_{{\cal D}_i} {\rm d}^3 q ~J_r =
\frac{1}{V_{\cal D}} \int_{{\cal D}_i} {\rm d}^3 q ~\dot{J_r}\nonumber \\
&=& \frac{1}{V_{\cal D}} \int_{{\cal D}_i} {\rm d}^3 q ~\theta J_r =
   \frac{1}{V_{\cal D}}
\int_{{\cal D}} {\rm d}^3 r ~\theta  =
\left < \theta \right >_{\cal D}
\label{eqn:local-H} \;,
\end{eqnarray}
where ${\rm d}/{\rm d}t$ denotes the Lagrangian time derivative.
Since $\left < \theta \right >_{\cal D} $ is the averaged volume expansion
rate of the domain ${\cal D}$, it may be natural to define the effective
Hubble expansion rate of the domain ${\cal D}$ by
$H_{\cal D}=\left < \theta \right >_{\cal D} /3$.
  From the definition of the volume $V_{\cal D}$, we find $H_{\cal
D}=\dot{a}_{\cal
D} / a_{\cal D} $.

The Lagrangian time derivative does not commute with spatial
averaging. For an arbitrary tensor field ${\cal A}$, Ehlers
and Buchert showed the commutation rule:
\begin{equation}
\frac{{\rm d} \left< {\cal A} \right >_{\cal D}}{\rm dt} -
\left < \frac{{\rm d} {\cal A}}{{\rm d} t} \right >_{\cal D} =
\left< \theta {\cal A} \right>_{\cal D} - \left < \theta
\right >_{\cal D} \left < {\cal A} \right >_{\cal D} \,.
\end{equation}

Applying this commutation rule to the basic equations, we find
\begin{eqnarray}
\frac{{\rm d}\left < \rho \right >_{\cal D}}{dt} &=& -\left < \theta \right
>_{\cal D} \left < \rho
\right >_{\cal D}  \;, \\
\frac{{\rm d}\left < \mbox{\boldmath{$\omega$}}\right >_{\cal D}}{dt}
&=& -\left <
\theta\right >_{\cal D}
\left < \mbox{\boldmath{$\omega$}}\right >_{\cal D}
+\left < \mbox{\boldmath{$\omega$}}\cdot \nabla_r {\bf u}
\right >_{\cal D}
   \;, \\
\frac{{\rm d}\left < \theta\right >_{\cal D}}{dt} &=& \Lambda -4\pi G
\left < \rho \right >_{\cal D}+\frac{2}{3}
\left < \theta^2 \right >_{\cal D}-\left < \theta \right>
_{\cal D}^2+2\left(\left < \omega^2
\right >_{\cal D}-\left < \sigma^2 \right >_{\cal D} \right)
\;. \label{eqn:theta_evolution}
\end{eqnarray}
Replacing $\left <
\theta\right >_{\cal D} $ in Eq. (\ref{eqn:theta_evolution}) with
$3\dot{a}_{\cal D}/a_{\cal D} $ , we find the averaged Raychaudhuri's
equation
\begin{equation}
3 \frac{\ddot{a}_{\cal D}}{a_{\cal D}} + 4 \pi G \left < \rho
\right >_{\cal D} - \Lambda = Q_{\cal D} \;,
\label{eqn:g-Friedmann}
\end{equation}
where
\begin{equation}
Q_{\cal D}\equiv \frac{2}{3} \left(\left < \theta^2 \right >_{\cal
D}-\left < \theta\right >_{\cal D}^2 \right)
+2\left(\left < \omega^2\right >_{\cal D}-\left <\sigma^2\right >_{\cal D}
\right)\;
\label{QD}
\end{equation}
describes the  backreaction due to inhomogeneity of the universe.
For $Q_{\cal D}=0$ case, the equation becomes the conventional
Friedmann equation with a scale factor $a(t)$ for
homogeneous and isotropic universe.
This is regarded as the generalized Friedmann equation.

\subsection{Evaluation of Backreaction Term $Q_{\cal D}$}

In order to evaluate the back reaction term $Q_{\cal D}$, it may be
convenient to introduce
   three principal scalar  invariants as follows:
For 2-rank tensor field $A=
(A_{ij})$~ in Cartesian coordinates, those  are defined by
\begin{eqnarray}
{\bf I}(A_{ij}) &=& {\rm tr}(A_{ij}) \;, \\
{\bf II}(A_{ij}) &=& \frac{1}{2}\left({\rm tr}(A_{ij})^2-
{\rm tr}
\left((A_{ij})^2\right)\right) \;, \\
{\bf III}(A_{ij}) &=& {\rm det}(A_{ij}) \;.
\end{eqnarray}
In particular,for the velocity gradient of matter fluid
~$(u_{i,j})$~, we find
\begin{eqnarray}
{\bf I}(u_{i,j})&=&u_{i,i}=\nabla_r\cdot {\bf u}=\theta
    \;, \label{eqn:append_Iu} \\
{\bf II}(u_{i,j})&=&\frac{1}{2}\left\{(u_{i,i})^2-
u_{i,j} u_{j.i}\right\} \nonumber \\
&=& \frac{1}{2}\nabla_r\cdot\left\{{\bf u} (\nabla_r \cdot
{\bf u})
-({\bf u} \cdot \nabla_r) {\bf u} \right\} \nonumber\\
&=& \omega^2-\sigma^2+\frac{1}{3}\theta^2 \;,
\label{eqn:append_IIu} \\
{\bf III}(u_{i,j}) &=& \frac{1}{3} u_{i,j}u_{j,k}u_{k,i}-
\frac{1}{2}(u_{i,i})
      (u_{j,k}u_{k,j})+\frac{1}{6}(u_{i,i})^3 \nonumber \\
    &=& \frac{1}{3} \nabla_r \cdot \left\{\frac{1}{2}\nabla_r
    \cdot
      \left({\bf u}(\nabla_r \cdot {\bf u})-({\bf u} \cdot
      \nabla_r)
      {\bf u} \right) {\bf u} \right. \nonumber \\
      && \qquad \left.-\left({\bf u} (\nabla_r\cdot {\bf u})
      -({\bf u} \cdot\nabla_r) {\bf u} \right)\cdot\nabla_r
      {\bf u}
      \right\} \nonumber \\
    &=& \frac{1}{9}\theta^3+2\theta\left(\sigma^2+\frac{1}{3}
    \omega^2
      \right)-\sigma_{ij}\sigma_{jk}\sigma_{ki}
      -\sigma_{ij}\omega_i\omega_j
\end{eqnarray}
where we have used the following relations
\begin{eqnarray}
\frac{1}{2} u_{i,j}u_{i,j}&=&\omega^2+\sigma^2+\frac{1}{6}
\theta^2 \;, \\
\frac{1}{2} u_{i,j}u_{j,i}&=&-\omega^2+\sigma^2+\frac{1}{6}
\theta^2 \;.
\end{eqnarray}

Using those invariants, we can describe  $Q_{\cal D}$ as
\begin{equation}
Q_{\cal D} =2\left < {\bf II}(u_{i,j}) \right >_{\cal D}-
\frac{2}{3} \left < {\bf I} (u_{i,j})\right >_{\cal D}^2 \;.
\label{eqn:QDinEuler}
\end{equation}

In the following discussion, we will use the Lagrangian description.
Then, it will be convenient to rewrite Eq. (\ref{eqn:QDinEuler}) by the
Lagrangian variables.
The Lagrangian domain is just the
initial domain ${\cal D}_i$. The spatial average of a tensor field $A$
   is
described  by the Lagrangian coordinates as
\begin{equation}
\left < A \right >_{\cal D} = \frac{1}{V_{\cal D}(t)}
\int_{{\cal D}} {\rm d}^3 r A=\frac{\int_{{\cal D}_i} {\rm d}^3 q J_rA}
{\int_{{\cal D}_i} {\rm d}^3 q J_r}
=\frac{\left < J_r A
\right >_{{\cal D}_i}}{\left < J_r \right >_{{\cal D}_i}}  \;,
\end{equation}
where the Lagrangian average $\left < A
\right >_{{\cal D}_i}$ is defined by
\begin{equation}
\left < A
\right >_{{\cal D}_i}={1\over V_{{\cal D}_i}}\int_{{\cal D}_i} {\rm d}^3 q
~ A
\end{equation}
Using this relation, the backreaction term is described  by the
volume average in the Lagrangian domain ${\cal D}_i$ as
\begin{equation}
Q_{\cal D}= \frac{2}{\left < J_r \right >_{{\cal D}_i}}
\left < J_r {\bf II}(u_{i,j}) \right >_{{\cal D}_i}-
\frac{2}{3} \left\{\frac{1}{\left < J_r \right >_{{\cal D}_i}}
\left <J_r {\bf I}(u_{i,j})\right >_{{\cal D}_i} \right\}^2
\;.
\label{eqn:QDinLagrange}
\end{equation}

So far, we have not made any approximation.
However, when we   evaluate those invariants explicitly,
we need a further ansatz.
Here we adopt a perturbation method.
In the conventional perturbation approach, we
expand the variables around the background Friedmann universe.
In fact, Buchert, Kerscher, and Sicka evaluated the backreaction term
$Q_{\cal D}$ by this perturbation method\cite{BucKerSic00}, which we will
show in the next section.
We have also another possibility to divide a perturbed part from a
uniform background part as follows:
Since we average the variables in some domain and take into account its
backreaction effect, we can adopt the averaged variables as unperturbed
parts. This way to extract the perturbed parts seems to be more
consistent and may provide some difference from the previous approach by
Buchert et al. We will discuss the detail next.

\section{Lagrangian perturbation equations with backreaction}
The conventional  Lagrangian approximation (for example, ZA
\cite{Zeldovich70}) was constructed in a homogeneous and
isotropic universe. However, when a nonlinear structure is
formed,
it is not so clear what is the  background universe and how to divide the
perturbed part from the unperturbed one.
Although the simplest way is the conventional approach, as we discussed
in previous section, a local expansion rate of the universe in some
finite domain will be affected by its inhomogeneity. Then the growth
rate of density fluctuations may also be affected by the inhomogeneity.

We shall derive the consistent  equations
for the  Lagrangian perturbations with the backreaction term $Q_{\cal D}$.
First, we introduce the  ``comoving"
Eulerian coordinates ${\bf x}_b$ and peculiar velocity ${\bf v}_b$ as
\begin{eqnarray}
{\bf r} &\equiv& a_b(t) {\bf x}_b \;, \\
{\bf u} &\equiv& \dot{\bf r} = \dot{a}_b(t) {\bf x}_b+{\bf v}_b \;.
\end{eqnarray}
In the conventional Lagrangian
perturbation approach, we set $a_b=a_H(t)$ where $a_H$ is just a  scale
factor of the background Hubble flow of the whole universe and satisfies
the Friedmann equation. However, because of the above reason, the
expansion of a domain ${\cal D}$ may be described by $a_{\cal D}$.
Hence, we have not fixed a scale factor
$a_b$ yet.
   It will be determined later by adding a further ansatz.
 
With the
relations
\begin{eqnarray}
\nabla_r &=& \frac{1}{a_b} \nabla_{{\bf x}_b} \;, \\
\left (\frac{\partial f}{\partial t}\right)_r &=&
-\frac{\dot{a}_b}{a_b}({\bf x}_b \cdot \nabla_{{\bf x}_b}) f
+\left(\frac{\partial f}{\partial t}\right)_{{\bf x}_b}  \;,
\end{eqnarray}
the basic equations (\ref{eqn:conti-eq})-(\ref{eqn:poisson-eq2}) with
$P=0$ are rewritten as
\begin{eqnarray}
\frac{\partial \delta_b}{\partial t} + \frac{1}{a_b} \nabla_{{\bf x}_b}
\cdot \{
{\bf v}_b (1+\delta_b) \} &=& 0  \label{eqn:Conti1} \;, \\
\left(\frac{\partial {\bf v}_b}{\partial t}\right)_{{\bf x}_b}+
\frac{1}{a_b}
({\bf v}_b \cdot\nabla_{{\bf x}_b}) {\bf v}_b+\frac{{\dot a}_b}{a_b} {\bf
v}_b &=& -\frac{1}{a_b} \nabla_{{\bf x}_b}
\left(\Phi+\frac{1}{2} a_b\ddot{a}_b x_b^2 \right) \;,
\label{eqn:Euler1} \\
\triangle_{{\bf x}_b}\left(\Phi-\frac{2}{3}\pi G \rho_b a_b^2 x_b^2 \right)
&=& 4 \pi G a_b^2 \rho_b \delta_b -\Lambda a_b^2\;, \label{eqn:Poisson1}
\end{eqnarray}
where
the density $\rho$ is divided into an unperturbed uniform part
$\rho_b(t) $, which is defined by the mass conservation $\rho_b a_b^3$=
constant, and its fluctuation
$\delta$ defined by
\begin{equation}
\delta_b \equiv  {\rho-\rho_b\over \rho_b }.
\end{equation}
   $\Phi$ is the gravitational potential defined by
${\bf g}= -\nabla_{\bf r} \Phi$.

  From Eqs. (\ref{eqn:Euler1}) and (\ref{eqn:Poisson1}),
we obtain the following equation:
\begin{equation}
\left(3\frac{\ddot{a}_b}{a_b}+4 \pi G\rho_b\right)
+\frac{1}{a_b}\nabla_{{\bf x}_b}\cdot\left\{\left(\frac{\partial {\bf
v}_b}{\partial t}\right)_{{\bf x}_b}
+\frac{1}{a_b}({\bf v}_b \cdot \nabla_{{\bf x}_b}){\bf v}_b
+\frac{\dot{a}_b}{a_b} {\bf v}_b
\right \}+4\pi G \rho_b \delta_b =0 \;.\label{eqn:evolution}
\end{equation}
If we specify $a_b$, this equation provides the equation for inhomogeneous
part subtracted a background uniform expansion.
For example, in the
conventional  Lagrangian approximation,
$a_b$ is taken to be a scale factor of the FRW universe $a_H$, which
satisfies the Friedmann equation
\begin{equation}
3\frac{\ddot{a}_H}{a_H}+4 \pi G \rho_H -\Lambda =0 \;,
\end{equation}
where $\rho_H$ is the mean energy density of the universe. On the other
hand, if
$a_b$ is chosen to be
$a_{\cal D}$, which evolution is given by  the averaged Raychaudhuri's
equation (\ref{eqn:g-Friedmann}), the perturbation equations are modified
   because of the  backreaction effect $Q_{\cal D}$ as
\begin{equation}
\frac{1}{a_{\cal D}} \nabla_{x_{\cal D}}\cdot\left\{\left(\frac{\partial
{\bf v}_{\cal D}}{\partial t}\right)_{x_{\cal D}}+\frac{1}{a_{\cal D}}
({\bf v}_{\cal D}\cdot\nabla_{x_{\cal D}}) {\bf v}_{\cal D}
+\frac{{\dot a}_{\cal D}}{a_{\cal D}} {\bf v}_{\cal D} \right
\}+4\pi G\rho_{\cal D}\delta_{\cal D} =-Q_{\cal D} \;,
\label{eqn:evolution2}
\end{equation}
where the variables with subscript ${\cal D}$ are defined those with the
scale factor $a_{\cal D}$ and $\rho_{\cal D}=\left<\rho\right>_{\cal D}$.

The rotation of (\ref{eqn:Euler1}) yields
\begin{equation}
\nabla_{{\bf x}_b}\times
\left\{\left(\frac{\partial {\bf v}_b}{\partial t}\right)_{{\bf x}_b}+
\frac{1}{a_b} ({\bf v}_b \cdot\nabla_{{\bf x}_b}) {\bf v}_b
+\frac{{\dot a}_b}{a_b} {\bf v}_b \right\}=0 \;.
\label{eqn:rot}
\end{equation}

Now we will describe those equations in terms of the  Lagrangian
coordinates. Using the Lagrangian time derivative along the fluid
flow
\begin{equation}
\frac{{\rm d}}{{\rm d} t} \equiv \frac{\partial}{\partial t}
+ \frac{1}{a_b}({\bf v}_b \cdot \nabla_{{\bf x}_b}) \,,
\end{equation}
Eqs. (\ref{eqn:evolution2}) and (\ref{eqn:rot}) are
rewritten as
\begin{eqnarray}
\frac{1}{a_b} \nabla_{{\bf x}_b} \cdot\left(\frac{{\rm d} {\bf v}_b}{{\rm
d} t}
+\frac{\dot{a}_b}{a_b} {\bf v}_b
\right)+4\pi G \rho_b \delta_b &=& -Q_b
\label{eqn:longitudinal} \;,\\
\nabla_{{\bf x}_b} \times\left(\frac{{\rm d} {\bf v}_b}{{\rm d} t}+
\frac{\dot{a}_b}{a_b} {\bf v}_b
\right)&=&0 \;,\label{eqn:transverse}
\end{eqnarray}
where we have introduced $Q_b$, which is zero for the conventional
Lagrangian approach without backreaction but is chosen to be $Q_{\cal D}$
when we include the backreaction due to inhomogeneity.
Note that there are still some space to estimate this backreaction term
depending on a choice of the background scale factor $a_b$.

The Lagrangian coordinates ${\bf q}_b$, which follows
the background flow, is defined by
the initial values of the comoving Eulerian coordinates ${\bf x}_b$,
and
the Lagrangian perturbations are described as
\begin{equation}
{\bf x}_b = {\bf q}_b + {\bf S}_b ({\bf q}_b, t) \;.
\label{eqn:x=q+s}
\end{equation}
where ${\bf S}_b$ denotes the displacement from the uniform distribution
assuming that the scale factor is given by $a_b$.
The continuity equation (\ref{eqn:Conti1}) or equivalently the mass
conservation yields
\begin{eqnarray}
\label{eqn:masscon}
dM=\rho d^3r =  \rho a_b^3 d^3x_b
= \rho a_b^3  J_b d^3q_b =\rho_i  d^3q_b={\rm constant},
\end{eqnarray}
where
$J_b \equiv {\rm det}  (\partial {\bf x}_{b i}/\partial {\bf q}_{b j})
   = {\rm det} (\delta_{ij} + \partial {\bf S}_{b i}/ \partial  {\bf q}_{b
j})$ is the Jacobian of the coordinate transformation ${\bf x}_b
\rightarrow {\bf q}_b$.
With $\rho_b a_b^3$= constant, we  have
\begin{equation}
\label{eqn:exactrho}
\rho \propto  \rho_b J_b^{-1} \,,
\end{equation}
or equivalently for density contrast
\begin{equation}
\label{eqn:exactdelta}
\delta_b = J_b^{-1} -1 \,.
\label{eq:delta}
\end{equation}

Now all physical quantities are found to be written in terms
of ${\bf S}_b$
and it remains only to find solutions for ${\bf S}_b$.
We obtain the following equations for ${\bf S}_b$ from
equations (\ref{eqn:longitudinal}) and
(\ref{eqn:transverse}):
\begin{equation}\label{eqn:poissonL}
\nabla_{{\bf x}_b} \cdot
\left(\ddot{\bf S}_b + 2\frac{\dot{a}_b}{a_b} \dot {\bf S}_b
   \right)
= -4\pi G \rho_b (J_b^{-1} -1) -Q_b\,.
\end{equation}
\begin{equation}\label{eqn:curlfreeL}
\nabla_{{\bf x}_b} \times
\left(\ddot{\bf S}_b + 2\frac{\dot{a}_b}{a_b}\dot{\bf S}_b \right) =0
\,,
\end{equation}

In what follows, the subscript $b$ will be dropped except for $a_b$  and
$\rho_b$, but we may  put it when we have to distinguish them.
In order to write Eqs. (\ref{eqn:poissonL}) and (\ref{eqn:curlfreeL}) in
terms of $ {\bf q}$ derivative, we use the relation between
$\nabla_{\bf x}$ and
$\nabla_{\bf q} \equiv \partial/\partial {\bf q}$, which is given  by the
definition as
\begin{equation}\label{eqn:partq-partx}
\frac{\partial}{\partial q_i} = \frac{\partial x_j}{\partial
q_i} \frac{\partial}{\partial x_j}
= \left(\delta_{ji}+\frac{\partial S_j}{\partial q_i} \right)
\frac{\partial}{\partial x_j}
= \frac{\partial}{\partial x_i}+\frac{\partial S_j}{\partial
q_i}  \frac{\partial}{\partial x_j}
\,.
\end{equation}
Using this equation iteratively, we have
\begin{eqnarray}
\frac{\partial}{\partial x_i} &=& \frac{\partial}{\partial
q_i}
-\frac{\partial S_j}{\partial q_i}\frac{\partial}{\partial
x_j} \nonumber \\
    &=& \frac{\partial}{\partial q_i} - \frac{\partial S_j}
    {\partial q_i}
\left(\frac{\partial}{\partial q_j}-\frac{\partial S_k}
{\partial q_j}
\frac{\partial}{\partial x_k}\right) \nonumber \\
    &=& \frac{\partial}{\partial q_i}
-\frac{\partial S_j}{\partial q_i}\frac{\partial}{\partial
q_j}
+\frac{\partial S_j}{\partial q_i}\frac{\partial S_k}
{\partial q_j}
   \frac{\partial}{\partial x_k} = \cdots \,.
\label{eqn:partx-partq}
\end{eqnarray}

This gives the differential equations in terms of ${\bf q}$ formally.
In order to write the basic equations explicitly, however,  we need
perturbative approach.
We then expand the deviation vector ${\bf S}$  as  ${\bf S}={\bf
S}^{(1)}+{\bf S}^{(2)}+\cdots$. Superscripts $(1)$ and $(2)$ denote
first  and  second order quantities in a perturbative expansion with
respect to
   magnitude of small displacement  $\epsilon$ from uniform distribution,
respectively. From Eq.  (\ref{QD}), the leading terms of the
backreaction is second order of $\epsilon$. Then we have to
derive the perturbation equations until  at least  second order.
  From Eq.
(\ref{eqn:transverse}), we obtain the first and second order equations as
follows:
\begin{eqnarray}
\nabla_q \times \left(\ddot{\bf S}^{(1)}+2 \frac{\dot{a}_b}{a_b}
\dot{\bf S}^{(1)}
\right) &=&0 \;, \label{eqn:firstT} \\
\left\{\nabla_q \times \left(\ddot{\bf S}^{(2)}+2\frac{
\dot{a}_b}{a_b} \dot{\bf S}^{(2)}
\right)\right\}_i&=&\varepsilon_{ijk} S^{(1)}_{l|j}\left(
\ddot{S}^{(1)}_{k|l}
+2\frac{\dot{a}_b}{a_b}\dot S^{(1)}_{k|l}\right) \;.
\label{eqn:secondT}
\end{eqnarray}

Next we consider equation (\ref{eqn:longitudinal}).
The Jacobian $J$ is expanded as
\begin{equation}
J = 1 + \nabla_q \cdot {\bf S}^{(1)}
        + \nabla_q \cdot {\bf S}^{(2)}
        + \frac{1}{2} \left[
    (\nabla_q \cdot {\bf S}^{(1)})^2
        - S_{i|j}^{(1)} S_{j|i}^{(1)} \right] + \cdots \,.
\label{eq:Jacobian}
\end{equation}
Thus we obtain the first order equation,
\begin{equation}
\nabla_q \cdot\left(\ddot{\bf S}^{(1)}+2\frac{\dot{a}_b}{a_b}
\dot {\bf S}^{(1)}
-4\pi G \rho_b {\bf S}^{(1)}\right)=0 \;, \label{eqn:first1}
\end{equation}
and the second order one,
\begin{eqnarray}
&& \nabla_q \cdot \left(\ddot{\bf S}^{(2)}
+2 \frac{\dot{a}_b}{a_b} \dot{\bf S}^{(2)}
-4\pi G \rho_b {\bf S}^{(2)}\right)  \nonumber \\
&& =S^{(1)}_{j|i}\left(\ddot{S}_{i|j}+2\frac{\dot{a}_b}{a_b}
S^{(1)}_{i|j}\right)
-2 \pi G \rho_b\left\{\left(S^{(1)}_{i|i}\right)^2+S^
{(1)}_{i|j}
S^{(1)}_{j|i}\right\}
-Q_b \;.
\label{eqn:second1}
\end{eqnarray}
In order to solve the above perturbation equations,
we  decompose ${\bf S}^{(1)}$ and ${\bf S}^{(2)}$
into the longitudinal and the transverse parts in the form
\begin{eqnarray}
{\bf S}^{(1)}&=&\nabla_q \psi+\mbox{\boldmath $\psi$}^T,\;
\nabla_q \cdot \mbox{\boldmath $\psi$}^T=0
\label{psi_def}\;,\\
{\bf S}^{(2)}&=&\nabla_q \zeta+\mbox{\boldmath $\zeta$}^T,
\quad \nabla_q \cdot \mbox{\boldmath $\zeta$}^T=0 \label{zeta_def}\;,
\end{eqnarray}
where $\psi$ and $\zeta$ are first-order and
second-order
scalar functions, respectively.
As for their physical meanings,
the first-order longitudinal and transverse parts are related
to linear density and vortical perturbations, respectively.
For the second-order level, however, such a simple
interpretation of the perturbation modes does not hold any more.

The first-order perturbation equations
(\ref{eqn:first1}) and (\ref{eqn:firstT})
   become
\begin{eqnarray}
\triangle_q \left(\ddot{\psi}+2\frac{\dot{a}_b}{a_b}\dot \psi-4 \pi
G \rho_b \psi
\right)=0 \;, \label{eqn:first2} \\
\nabla_q \times \left(\ddot{\mbox{\boldmath $\psi$}}^T+2
\frac{\dot{a}_b}{a_b}
\dot{\mbox{\boldmath $\psi$}}^T \right)=0 \;.
\end{eqnarray}
The second-order perturbation equations
(\ref{eqn:second1}) and (\ref{eqn:secondT})   are also
divided into the longitudinal and the transverse parts as
\begin{eqnarray}
&& \triangle_q \left(\ddot{\zeta}+2\frac{\dot{a}_b}{a_b} \dot{
\zeta}-4
\pi G\rho_b \zeta\right) \nonumber \\
&=& 2 \pi G \rho_b \left\{\psi_{|ij}\psi_{|ij}-(\triangle_q
\psi)^2- \mbox{\boldmath $\psi$}^T_{i|j}
\mbox{\boldmath $\psi$}^T_{j|i}\right \}-Q_b
\label{eqn:second_long}\;,
\end{eqnarray}
\begin{equation}
\left[\triangle_q \times \left(\ddot{\mbox{\boldmath $\zeta$}}^T+2
\frac{\dot{a}_b}{a_b}
\dot{\mbox{\boldmath $\zeta$}}^T\right)\right]_i=4 \pi G \rho_b
\varepsilon_{ijk}
\mbox{\boldmath $\psi$}^T_{l|j} \psi_{|kl}
\label{eqn:second_trans}\;.
\end{equation}

In what follows, we will fix $a_b$ to be $a_{\cal D}$.
Note that if we set $a_b=a_H$, the above equations are equivalent
to those by Buchert et al.
Since the displacement vector
${\bf S}$ denotes a deviation from homogeneous distribution in the
local domain ${\cal D}$, the average density $\rho_b= \left < \rho \right
>_{\cal D}$, which is written by  $\delta_{\cal D}$ in what follows,  is
not  equal to that in the whole universe.
The density fluctuation described by
${\bf S}$ should be  defined as
\begin{eqnarray}
\delta_{\cal D} = \frac{\rho - \rho_{\cal D}}{\rho_{\cal
D}} ~~~~
{\rm with}~~~~~\left < \delta_{\cal D} \right >_{\cal D} = 0
\label{eqn:local-delta1} \;.
\end{eqnarray}
Here we use the subscript ${\cal D}$ to show what is the background mean
density.

   To solve the above equations, we carry out further
procedure. Since the first order equations (\ref{eqn:first2}) are the
same as those of ZA except for a scale factor, we can easily solve them.
The perturbation variables are separated into time and spatial
functions as
\begin{equation}
\psi_{\cal D} = b_{\cal D} (t) \varphi_{\cal D} ({\bf q}) \;.
\label{eqn:1st-psi}
\end{equation}
The evolution equation for $b_{\cal D}$ is
\begin{equation}
\ddot{b}_{\cal D} + 2 \frac{\dot{a}_{\cal D}}{a_{\cal D}}
\dot{b}_{\cal D} - 4 \pi G \rho_{\cal D} b_{\cal D} = 0 \;,
\label{eqn:bD}
\end{equation}
while $\varphi_{\cal D} ({\bf q})$ is determined from the initial data.

As for the second order equations, we need a little more complicated
procedure.
In Appendix A, we rewrite the backreaction term $Q_{\cal D}$ in context of
the Lagrangian approximation and present its explicit form up to  the
second order perturbations ($\psi$,$\mbox{\boldmath $\psi$}^T$,$\zeta$,
and $\mbox{\boldmath
$\zeta$}^T$).
   Since the transverse modes ($\mbox{\boldmath $\psi$}^T$
and $\mbox{\boldmath
$\zeta$}^T$) may not be
so important, we shall  ignore  them in what follows.
Then the  leading order of the  backreaction term $Q_{\cal D}$ becomes
very simple   as
\begin{equation}
Q_{\cal D} = 2 \left < {\bf II} (\dot{\psi}_{{\cal D}|ij})
\right > - \left < {\bf I} (\dot{\psi}_{{\cal D}|ij}) \right>
^2 \;,
\end{equation}
where $|i= \partial_{{\bf q}_i}$ (see Eq. (\ref{eqn:LA_QD_2nd}).
   Using (\ref{eqn:1st-psi}), we find
\begin{equation}
Q_{\cal D}=\frac{2}{3}\dot b_{\cal D}^2\left\{\frac{3}{2}
\left < (\triangle_{q_{\cal D}}
\varphi_{\cal D})^2
-\varphi_{{\cal D}|ij}\varphi_{{\cal D}|ji}\right >_{{\cal
D}_i}
-\left < \triangle_{q_{\cal D}}\varphi_{\cal D} \right >_{{
\cal D}_i}^2\right\} \;,
\label{eqn:backreaction}
\end{equation}
which depends only on time $t$.
Eq. (\ref{eqn:second_long}) for $\zeta_{\cal D}$ is now
\begin{eqnarray}
\triangle_q \left(\ddot{\zeta}_{\cal D}+2\frac{\dot{a}_b}{a_b}
\dot{\zeta}_{\cal D}-4
\pi G\rho_b \zeta_{\cal D}\right)
=2 \pi G \rho_{\cal D}
\left(\psi_{|ij}\psi_{|ij}-(\triangle_q
\psi)^2 \right) -Q_b \;,
\label{eqn:second-long}
\end{eqnarray}
Since the
first term of r.h.s. of Eq. (\ref{eqn:second_long})
\begin{equation}
2 \pi G \rho_{\cal D} b_{\cal D}^2  \left\{\varphi_{{\cal D}|
ij} \varphi_{{\cal D}|ij}-(\triangle_q \varphi_{\cal D})^2
\right\} \;
   \label{eqn:phi2}
\end{equation}
   depends on spatial coordinate ${\bf q}$ as well as $t$, we have to divide
a solution into two parts: one is inhomogeneous term and the other is
homogeneous one. First, in order to find a homogeneous term,
   we commute the time and spacial derivatives in Eq.
(\ref{eqn:second-long}), and then integrate it  over ${\cal D}_i$,
finding
\begin{eqnarray}
&& {\frac{{\rm d}^2}{{\rm d} t^2} \left < \triangle_{q_{\cal
D}} \zeta_{\cal D} \right >_{{\cal D}_i}
+2\frac{\dot{a}_{\cal D}}{a_{\cal D}}\frac{\rm d}{{\rm d} t}
\left <
\triangle_{q_{\cal D}}
\zeta_{\cal D} \right >_{{\cal D}_i}
-4\pi G \rho_{\cal D} \left < \triangle_{q_{\cal D}} \zeta_{
\cal D} \right >_{{\cal D}_i}} \nonumber \\
&= & 2 \pi G \rho_{\cal D} b_{\cal D}^2 \left < \varphi_{{
\cal D}|ij}\varphi_{{\cal D}|ij}
-(\triangle_{q_{\cal D}} \varphi_{\cal D})^2 \right >_{{\cal
D}_i}
-Q_{\cal D} \;,
\label{eqn:average-zetaD}
\end{eqnarray}
which is an ordinary differential equation for
$\left < \triangle_{q_{\cal
D}} \zeta_{\cal D} \right >_{{\cal D}_i}$.
By Introduction of  new variable $\tilde{\zeta}_{\cal D}$ by
\begin{equation}
\triangle_{q_{\cal D}}\tilde{\zeta}_{\cal D} \equiv
\triangle_{q_{\cal D}}\zeta_{\cal D}-
\left < \triangle_{q_{\cal D}}\zeta_{\cal D} \right >_{{\cal
D}_i} \;, \label{eqn:coord2}
\end{equation}
we can eliminate the backreaction term $Q_{\cal D}$.  The
equation for $\tilde{\zeta}_{\cal D}$ is now separable by setting
\begin{equation}
\tilde{\zeta}_{\cal D} = c_{\cal D} (t) \chi_{\cal D}({\bf
q}) \;,
\end{equation}
as
\begin{eqnarray}
&&\ddot{c}_{\cal D}+2\frac{\dot{a}_{\cal D}}{a_{\cal D}}
\dot{c}_{\cal D}
-4 \pi G \rho_{\cal D} (c_{\cal D}+A b_{\cal D}^2) =0 \;,
\label{eqn:cD} \\
&& \triangle_{q_{\cal D}} \chi_{\cal D}= \frac{\left < {\bf
II} (\varphi_{{\cal D}|ij}) \right >_{{\cal D}_i}-{\bf II}
(\varphi_{{\cal D}|ij})}{A} \;,
\label{eqn:2nd_init_condition}
\end{eqnarray}
where $A$ is a separation constant.
Finally  we obtain a set of basic equations
(\ref{eqn:g-Friedmann}),(\ref{eqn:bD}), (\ref{eqn:average-zetaD}), and
(\ref{eqn:cD}) with  the definition (\ref{eqn:backreaction}) for
dynamical variables
$a_{\cal D}$, $b_{\cal D}$,
$\left < \triangle_{q_{\cal
D}} \zeta_{\cal D} \right >_{{\cal D}_i}$, and $c_{\cal D}$.
The spatial inhomogeneities
$\varphi_{\cal D}({\bf q})$ and $\chi_{\cal D}({\bf q})$ are determined
by initial
distributions and Eq. (\ref{eqn:2nd_init_condition}).

\section{Relation between two Lagrangian descriptions \\: Setting up
Initial Data} In order to solve our basic equations for
Lagrangian perturbations, we have to set up our initial data.
Since the initial fluctuation is given by deviation from uniform
distribution of the whole universe, which is described by the Eulerian
coordinates, we have to construct our initial data from those
data by an appropriate  transformation.
   Furthermore, since our equations are valid only in each
domain
${\cal D}$,  if we wish to analyze some statistical properties of our
results, we have to go back to  the whole universe. Here we
will first discuss such a transformation, and then set up our initial data.

Suppose that we have some inhomogeneous distribution in the Eulerian
coordinates ${\bf r}$.  We then find $a_H$ and $a_{\cal D}$ by integrating
a volume over the whole universe and over a domain ${\cal D}$,
respectively. Here we set these initial values equal to unity. These scale
factors fix
${\bf x}_H$ (then
${\bf q}_H$ and
${\bf S}_H$), and ${\bf x}_{\cal D}$ (then ${\bf q}_{\cal D}$ and
${\bf S}_{\cal D}$). We also find $\rho_H$ and $\rho_{\cal D}$ by
averaging the density $\rho({\bf r}, t)$  over the whole universe and over
the domain ${\cal D}$, respectively.

The density perturbation in our basic equations is  deviation from
a density averaged in a domain ${\cal D}$. Hence a perturbation in
the conventional Lagrangian approximation does not satisfy the condition
(\ref{eqn:local-delta1}). We have to reconstruct
initial conditions in our local domain  from fluctuations given in the
whole universe.
In order to set up initial conditions, we may need only the relation
between the Eulerian coordinates ${\bf r}$ and our Lagrangian
coordinate ${\bf q}_{\cal D}$.
However, we also need the relation between two Lagrangian coordinates,
${\bf q}_{\cal D}$ and ${\bf q}_H$ from the following reason.
In an inhomogeneous universe, we do not know in which domain we are
living. Then, when we analyze our results, we need  a statistical analysis
in the whole universe to know how the observed values are plausible. In
particular, one of the most interesting observed values is  a peculiar
velocity, for which we have to introduce a Hubble flow $a_H$ and the
comoving Eulerian coordinate
${\bf x}_H$.   As a result,  we need a relation between ${\bf q}_H$ and
${\bf q}_{\cal D}$ as well. Note that although both Lagrangian coordinates
follow matter fluid, those  scales are different. Because we have
mass conservation for infinitely small domain (${\bf q}_{\cal D},{\bf
q}_{\cal D}+\Delta{\bf q}_{\cal D})
$, or equivalently  (${\bf q}_H,{\bf
q}_H+\Delta{\bf
q}_H) $, i.e.
\begin{eqnarray}
\Delta M=\rho_H a_H^3 (\Delta q_H)^3 = \rho_{\cal D} a_{\cal D}^3 (\Delta
q_{\cal D})^3 .
\end{eqnarray}
Assuming
\begin{equation}
a_{\cal D} (t_i) = a_H(t_i) =1
\label{eqn:initial_a} \;,
\end{equation}
we find
\begin{equation}
\frac{\Delta q_{\cal D}}{\Delta q_H} = \left[\frac{\rho_H(t_i)}{
\rho_{\cal D}(t_i)}\right]^{1/3}  \;.
\label{eqn:qd-qh1}
\end{equation}
Since the density perturbation in the conventional Lagrangian
approximation is given by
\begin{eqnarray}
\delta_H ={\rho-\rho_H\over \rho_H},
\end{eqnarray}
   we find $\rho_{\cal
D}=\rho_H(1+\left<\delta_H\right>_{\cal D})$ from the definition of
$\rho_{\cal D}$. This gives the relation between $q_{\cal D}$ and $q_H$
as
\begin{equation}
\frac{\Delta q_{\cal D}}{\Delta q_H} =
(1+\left<\delta_H(t_i)\right>_{{\cal
D}_i})^{-{1\over 3}}  \;. \label{eqn:qd-qh2}
\end{equation}
We then find the relation between
$\delta_H$ and
$\delta_{\cal D}$ as follows:
From
\begin{eqnarray}
   1+\delta_{\cal D} ={\rho \over \rho_{\cal D}} = {\rho \over
\rho_H}{\rho_H \over \rho_{\cal D}}= (1+\delta_H) {\rho_H (t_i)\over
\rho_{\cal D}(t_i)}{ a_{\cal
D}^3(t)
\over
   a_H^3(t)}
\end{eqnarray}
with Eq.(\ref{eqn:qd-qh2}), we find
\begin{eqnarray}
\delta_{\cal D} = \frac{a_{\cal D}^3
   ( 1+ \delta_H ) }{a_H^3\left [ 1 + \left < \delta_H (t_i)
\right >_{{\cal D}_i} \right ]} -1 \;,
\label{eqn:convert-delta}
\end{eqnarray}
We also obtain a peculiar
velocity against the Hubble
flow  as
\begin{eqnarray}
{\bf v}_H \equiv \dot{{\bf r}}- H{\bf r} =a_H \dot{\bf
S}_H =
\left(\dot{a}_{\cal D} -H a_{\cal D} \right){\bf x}_{\cal
D} + a_{\cal D}
\dot{\bf S}_{\cal D}
\;.
\label{eqn:convert-v}
\end{eqnarray}
We will use this definition in analysis of a peculiar velocity.

Now we can set up our initial condition.
As for the density perturbations, we find from
Eq. (\ref{eqn:convert-delta}) as
\begin{eqnarray}
\delta_{\cal D}(t_i) = \frac{
   ( 1+ \delta_H(t_i) ) }{\left [ 1 + \left < \delta_H (t_i)
\right >_{{\cal D}_i} \right ]} -1 \;.
\label{eqn:initial_delta}
\end{eqnarray}
  From this equation with definitions (\ref{eq:delta}), (\ref{eq:Jacobian}),
(\ref{psi_def}), and  (\ref{eqn:1st-psi}),
we find  the initial condition
for
$\varphi_{\cal D}$ as
\begin{equation}
\triangle_q \varphi_{\cal D} = \frac{\left < \delta_H (t_i)
\right >_{{\cal D}_i} - \delta_H (t_i)}{1 + \delta_H (t_i)}
\;,
\end{equation}
where we set $b_{\cal D}(t_i)=1$.

As for the
scale factor $a_{\cal D}$ of the domain ${\cal D}$,
its initial value is set to be unity (Eq.(\ref{eqn:initial_a})),
but its time derivative is non-trivial. In fact,
  From (\ref{eqn:local-H}), we find $\dot{a}_{\cal D}$ as
\begin{equation}
\dot{a}_{\cal D} (t_i) = \dot{a}_H (t_i) \left[ 1+ \frac{1}{3}
\left < {\bf I} \left (\psi_{{\cal D}, ij} \right ) \right >
_{{\cal D}_i} \right] \;. \label{eqn:aD-init}
\end{equation}

\section{Effect of inhomogeneity}
Here, we study a simple
model to show new aspect in our approach. We assume
a plane-symmetric 1-dimensional model.
The Lagrangian perturbation is
given by
\begin{equation}
\nabla_q \psi = {\bf S} ({\bf q}) = b(t) (s(q_1), 0, 0) \;.
\end{equation}
In the conventional Lagrangian approximation, ZA gives an exact solution
in a plane-symmetric case.
However, it does not take into account a backreaction effect of
inhomogeneity on the Hubble expansion.
Since our approach includes the backreaction effect,
we will analyze this simple one-dimensional model and compare our results
with those by ZA.
We also look at a difference of the backreaction term estimated by the
conventional Lagrangian approach (Buchert et al).

   When we consider
our backreaction effect, we have a set of  equations:
\begin{eqnarray}
&& \ddot{\psi}_{\cal D} + 2\frac{\dot{a}_{\cal D}}{a_{\cal D}}
\dot{\psi}_{\cal D} + 4 \pi G
\rho_{\cal D}
\psi_{\cal D} = -Q_{\cal D} \,, \\
&& Q_{\cal D}= \frac{2}{3} \frac{\left ( \dot{b}_{\cal D}^2 \left < {
\bf I}_s \right >_{{\cal D}_i} \right )^2}{\left (1 + b^2
\left < {\bf I}_s \right >_{{\cal D}_i} \right )^2} \,.
\end{eqnarray}

Since we analyze the present model numerically, we have to introduce the
size of "whole universe", which is $L$. We the introduce the scale of
averaging domain $l$.
As for initial
conditions, we adopt a power law spectrum with the index $n=1$:
\begin{equation}
P_i(k) \propto k \;.
\end{equation}
We also introduce a  cutoff at small scale, which wave number is
$k_{cut} = 1024 k_0$, where $k_0 = 2\pi/L$.
We set that the initial time is $a=1$.
Then the amplitude of fluctuation is
chosen so that
the first
shell-crossing occurs at $a \simeq 1000$.
   The number
of grids is  $N=2^{16}$ and we use a periodic boundary condition.
We take an ensemble average over 500 samples, which initial
conditions are given by random Gaussian.
In our approach, since we do not know where we are living, we study its
statistical properties. In particular, we will see
the scale dependence of the averaged variables and the
probability distribution of the Hubble parameter, deceleration parameter,
and pair-wise velocity.

\subsection{Hubble parameter}
First we analyze the expansion rate of local domain. If we
fix the  Hubble parameter $H_0$ by local observation in a domain ${\cal
D}$, the most probable value of  $H_0$ is given by the averaged expansion
rate of the domain, which is $\left<\theta\right>_{\cal D}/3$. Several
authors so far discussed such a local measurement of the Hubble parameter
\cite{TurCenOst92,SutSugIna95,NakSut95,Wu95,ShiTur98}.
In particular,
Shi and Turner \cite{ShiTur98} estimated a possible value of
the Hubble constant measured locally  and discussed a deviation from
the global value, using linear perturbation theory with the CDM model.
They found that for small
samples of objects that only extend to 10,000 km $s^{-1}$, the variance can 
reach
4\%, while for large samples of objects to 40,000 km $s^{-1}$,
the variance is about 1-2 \%.

\begin{figure}
\begin{center}
\epsfig{figure=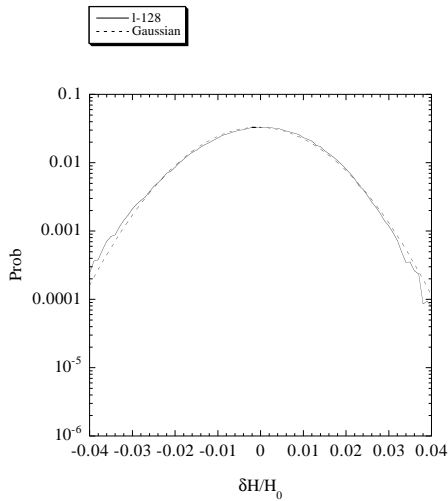,width=6cm}
\end{center}
\caption{\label{fig:PDF-H0} The PDF of the Hubble
parameter at $a=900$ for $l=128$-grid domain. The dispersion is  1.2 \%.}
\end{figure}

We solve Eq. (\ref{eqn:g-Friedmann}) for $a_{\cal D}$ with a backreaction
due to inhomogeneity. If we are living in an underdense region on average,
the expansion rate will  be faster than the Hubble one for the whole
universe. While, if we stay in an overdense region, the rate
will be slower than the global Hubble one. Fig.\ref{fig:PDF-H0} shows
the PDF of a local Hubble parameter in our calculation. If our domain is
   small, the deviation from $H_0$ gets large. For example, the
dispersion of the Hubble parameter is about 1.2 \% for
the $l=128$-grid domain, while 0.66 \% for the$l=256$-grid domain.
  The dispersion of our model is consistent with the result
by Shi and Turner.

\subsection{Density fluctuation}
Next we show the PDF of density fluctuations. In the Eulerian
linear approximation, if initial data is given by random
Gaussian distribution, the PDF of density fluctuations will remain its
Gaussian form during evolution. On the other hand, in the Lagrangian
approximation, there appears a nonlinear effect. In fact, Kofman et al
   shows that the PDF approaches to a log-normal function
rather than a Gaussian function in the cases of the Lagrangian
approximation and N-body simulation\cite{Kofman94}. Padmanabhan and
Subramanian
   also discussed the PDF with the ZA and found a non-Gaussian
distribution\cite{PadSub93}.

\begin{figure}
\begin{center}
\epsfig{figure=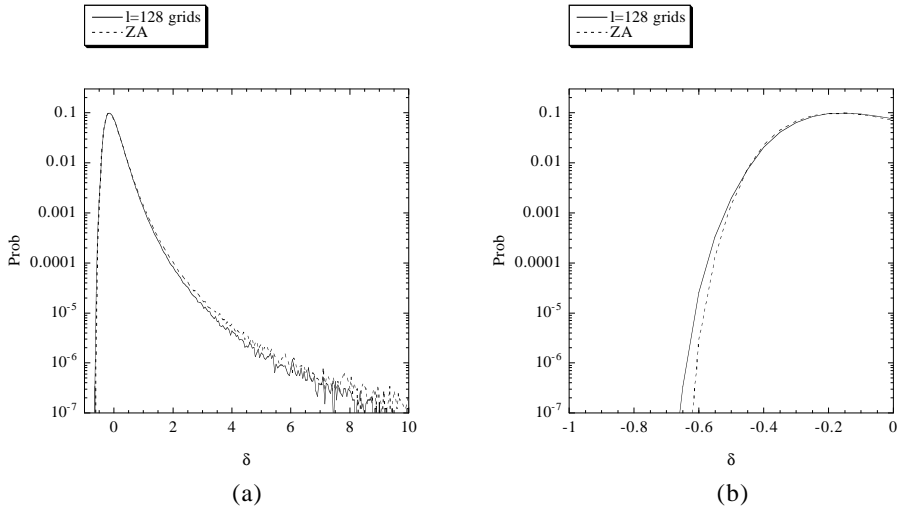,width=12cm}
\end{center}
\caption{\label{fig:PDF-delta} The PDF of density fluctuation
at $a=900$ for $l=128$-grid domain and that with the ZA. The
probability to find an  overdense region for $l=128$-grid domain is
less than that with the ZA((a)). On the other hand, the
probability for an underdense region increases for $l=128$-grid
domain((b)).}
\end{figure}

Here we analyze the PDF of density fluctuations using our approximation.
The results are shown in Fig.\ref{fig:PDF-delta}. From comparison
with the result of ZA, the
void region (i.e. an underdense region; $\delta <0$) is found
in higher probability in our  approximation. Especially, if the size of a
domain  is smaller, the difference gets larger.
On the other hand, the probability to find an overdense region ($\delta
>0$) decreases in our approximation.

The reason is very simple: During evolution, an overdense region
shrinks and a nonlinear structure is formed  as the Zel'dovich's
pancake. On the other hand, an underdense region expands.
Therefore, although the initial volumes of overdense
underdense regions are  the same, the volume of the latter
gets larger than that of the former in a nonlinear stage. In addition
to this Lagrangian nonlinear effect, we take into account a  backreaction
effect. This effect enhances expansion of an underdense region
and  contraction of an overdense region. As a result, the above difference
between ZA and our approximation appears.

\subsection{Peculiar velocity distribution}
\begin{figure}
\begin{center}
\epsfig{figure=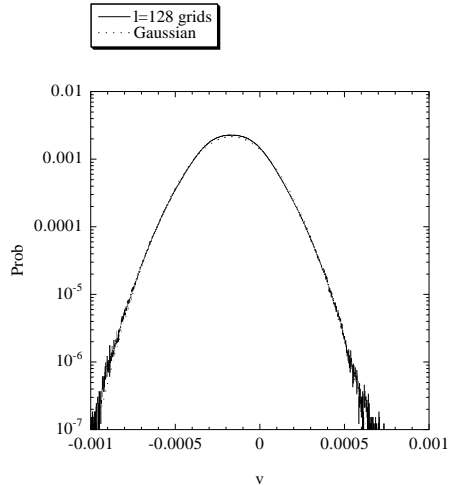,width=6cm}
\end{center}
\caption{\label{fig:PDF-v} The PDF of peculiar velocity at
$a=900$ for $l=128$ grid domain. The PDF obeys Gaussian form.}
\end{figure}

In the conventional Lagrangian approximation,
if an initial condition is given by
a random Gaussian distribution, the PDF of a peculiar velocity also remains
its Gaussian form\cite{Kofman94}.
Because a peculiar velocity ${\bf v}_H$ in the conventional Lagrangian
approximation (or the ZA) is given by
\begin{equation}
{\bf v}_H = a_H\dot{\bf S}_H \;,
\end{equation}
the spectrum of a peculiar velocity is proportional to that of a density
fluctuation as
$P_v(k) \propto (\dot{b}/B)^2 P_\delta (k)$.
   However, in our case, a peculiar velocity is given by
(\ref{eqn:convert-v}) and  the growing
factor $b_{\cal D}$ is different in each domain, the PDF could deviate from
a Gaussian distribution.
However, from our numerical analysis (see Fig.
\ref{fig:PDF-v}), the  PDF of a peculiar velocity
seems to still be a Gaussian.
This may be because we need only a linear
term of the Lagrangian perturbations in the present analysis.

\subsection{Pairwise peculiar velocity distribution}
The PDF of a radial pairwise peculiar velocity is known to show
an exponential form from analysis of  N-body simulation and the ZA
\cite{ZQSW94,SetYok98}. For this variable, even if initial data is
given by a random Gaussian distribution,
the PDF approaches an exponential
form as the universe evolves\cite{SetYok98}.
The origin of this result could be understood by  nonlinearity of gravity.
In the case of one-dimensional plane-symmetric system in the conventional
Lagrangian approximation (in fact, the ZA is exact), we do not find any
non-Gaussian structure. However, even in the ZA, the PDF shows
non-Gaussian behavior in the case of three-dimensional
case\cite{SetYok98}.  If we
take into account  our backreaction effect, does non-Gaussian behavior
emerge even in one-dimensional case ?

The pairwise peculiar velocity is defined as follows:
\begin{eqnarray}
{\bf v}_{AB} (t) &\equiv& {\bf v}_B (t) - {\bf v}_A (t)
\nonumber \\
&\equiv& {\bf v}_{\parallel} (t) + {\bf v}_{\perp} \;,
\label{eqn:Def-PV}
\end{eqnarray}
where ${\bf v}_{\parallel}$ and ${\bf v}_{\perp}$ represent
components parallel and perpendicular to ${\bf x}_{AB} \equiv
{\bf x}_B - {\bf x}_A$, respectively. In a plane-symmetric case,
   ${\bf v}_{
\parallel}$ only appears. Hereafter we write this by
$\tilde{v}$. If matter distribution is clustering,
$\tilde{v}$ is expected to be
negative.

\begin{figure}
\begin{center}
\epsfig{figure=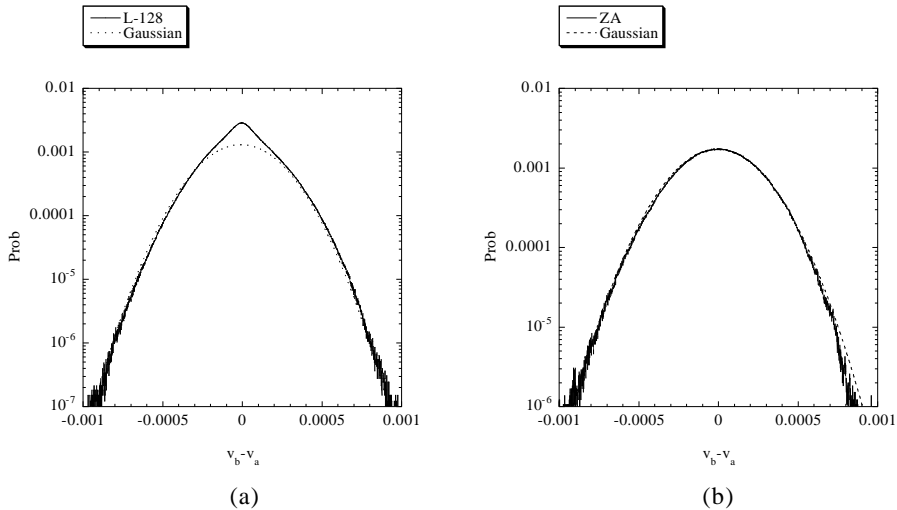,width=12cm}
\end{center}
\caption{\label{fig:PDF-pv} The PDF of pairwise peculiar
velocity at $a=900$ for $l=128$-grid domain (a) and E-dS model
(b). The PDF in (a) does not obey Gaussian form, and
approaches to log-normal form at small velocity region. On
the other hand, the PDF in (b) obeys Gaussian form.}
\end{figure}

Giving initial data by random Gaussian, the PDF of a pairwise
peculiar velocity is Gaussian at initial time. During
evolution, the PDF will deviate from Gaussian.
In fact, from Fig.\ref{fig:PDF-pv}, which
shows the PDF of a peculiar velocity in nonlinear regime,  we
find that it is not Gaussian
and approaches an exponential form in a small velocity region.
As a reference, we show that the PDF for the ZA, which shows a Gaussian
form. The reason may be understood as follows:
   In a plane-symmetric model, gravitational
potential is proportional to a distance between two sheets
($\psi \sim |r|$). Then, even if two sheets approaches
very closely, a gravitational force does not become strong but keep
constant. When we take into account a backreaction effect, however, a
gravitational force will be  strong in a clustering region, because the
expansion rate of a local domain is slow down. The
strengthening of a gravitational force in a cluster region may make a
deviation of the PDF of pairwise peculiar velocity from its Gaussian
form. Note that in the 3D system, which shows non-Gaussian PDF, a
gravitational force increases when two particles approach.

\subsection{Deceleration parameter $q$}

Another interesting observable variable is a deceleration parameter.
The recent observation of type Ia supernova may suggest an acceleration of the
Universe\cite{Perlmutter99}.
Although this  result may naively suggest an existence of dark energy such as a
cosmological constant $\Lambda$, we could find some effective model without
dark energy which explain the observation.
Then we shall estimate  a deceleration parameter averaged in a local domain 
here.

We define local deceleration parameter $q_{\cal D}$ as
\begin{equation}
q_{\cal D} \equiv  - \frac{a_{\cal D} \ddot{a}_{\cal D}}{\dot{a}_{\cal 
D}^2} \;,
\end{equation}
which can be evaluated by Eq. (\ref{eqn:g-Friedmann}) and $H_{\cal D}$.
Buchert et al\cite{BucKerSic00} showed the evolution of deceleration
parameter for the ZA. They picked up overdense and underdense regions of
three-$\sigma$ fluctuations and found that  $q_{\cal D}$ for an underdense
region could be a present day value, which is  smaller  by more than 200\% than
that of the background E-dS Universe, although such a region is still
decelerating.

\begin{figure}
\begin{center}
\epsfig{figure=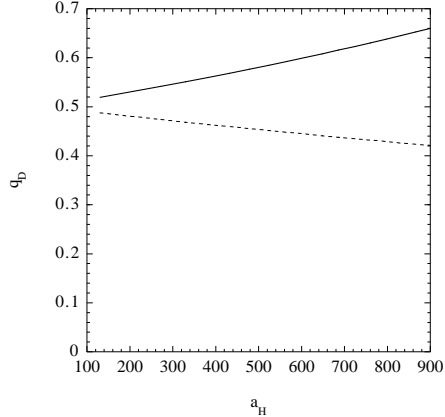,width=6cm}
\end{center}
\caption{\label{fig:qD-evolv} The evolution of a deceleration parameter
$q_{\cal D}$ for $l=64$-grid domain. The overdense domain shows increase of
$q_{\cal D}$ (a solid line for maximum value),  while the underdense domain
shows decrease of
$q_{\cal D}$ (a dashed line for minimum value). }
\end{figure}

Although our approach includes a backreaction consistently, our analysis shows
that a deviation of
$q_{\cal D}$ does not get so large. The difference of $q_{\cal D}$ from the  ZA
is very  little even just before the shell crossing. We show the time
evolution of $q_{\cal D}$ for a plane-symmetric 1-dimensional  model in Fig.
\ref{fig:qD-evolv}.   Even if a domain is extremely underdense, the
domain is decelerating. This may be because our approach is still perturbative.
We will discuss it further in the next section.

\section{Conclusion and Remarks}
We propose new Lagrangian perturbation theory with a backreaction
effect by inhomogeneity of density perturbations and present a set of
basic equations.
   The inhomogeneity affects
the expansion rate in a local domain and its own growing rate.
In a one-dimensional  plane-symmetric
model,  we have numerically analyzed our approach, and
calculated the growing rate density perturbations and the PDF of several
observed variables.
We set our initial conditions as random Gaussian distribution.
  From our analysis, we show that  the expansion
rate of an overdense region is faster than that of the whole universe
as expected. We also show that the local Hubble
parameter may deviate from the global one by  about 1.2 \% for
a $l=128$-grid domain. It
may be too small to distinguish its effect in the
present observations\cite{Freedman01}, but it could become important in
fully non-linear stage, which we cannot describe in the present approach.

The PDF of
density is slightly different from that of the ZA. In our model,
an underdense region expands faster than that in the E-dS  model and then
its volume gets larger.
Hence, a probability for a  negative $\delta$ region increases as seen in
the PDF.
As for a peculiar velocity,
even if we take into account a backreaction, its PDF
is still  Gaussian.

The PDF of
pairwise peculiar velocity, however,  shows an effective difference
from the conventional Lagrangian approach.
In one-dimensional plane
symmetric case, the PDF in the conventional Lagrangian approximation
(the ZA) is  Gaussian, but ours is not but approaches  an
exponential form in a small relative-velocity region, which agree with
the N-body simulation and the 3D Lagrangian
approximation\cite{ZQSW94,SetYok98}.

Finally, we mention about recent observation about cosmological
parameters. According to the observation of type Ia supernova,
the expansion  of the Universe seems to accelerate\cite{Perlmutter99}.
Combining observation of the cosmic microwave background radiation (CMBR),
the result suggests existence of dark energy such as a cosmological
constant
$\Lambda$\cite{Jaffe01}.
However, this produces another difficulty, that is the so-called
cosmological constant problem.
To avoid such a difficulty, if we could explain the observation without
cosmological constant, it would be more natural.
Recently, Tomita discussed such possibility assuming we are in a large
local void\cite{Tomita00a,Tomita00b,Tomita01}. Globally the Universe is
flat (EdS universe), but we are sitting near the center of a local void,
which existence is observationally confirmed. Then he
calculated the luminosity distance, finding that the observation can be
explain by such a model.
We then wonder whether we could have the similar explanation if we are
living in an effective void, which is a domain with an averaged energy
density below the critical value.
Since we have to treat a strongly nonlinear structure to explain the
observation, this is beyond our present approach.
However there is some indication.
If we analyze the time dependence of the backreaction term $Q_{\cal D}$,
which evolves as $a^{-1}$ in a linear perturbation level.
This time dependence is the same as a perfect fluid with the equation of
state $P=-{2\over 3}\rho$, which could be dark energy.
   If we could explain the above observation without introduction of any
strange matter but just by  inhomogeneity of density distribution in the
Universe,  we will have a natural understanding of the Universe.
This is under investigation.

In addition, there are two further approaches which are related to the
present work. Takada and Futamase\cite{TakFut99} proposed that they
divided Lagrangian perturbation to large-scale and small-scale
perturbations, then discussed interaction between those scales.
Taruya and Soda\cite{TarSod00}
discussed dynamics of averaged variables in the case of a  spherical infall
model taking into account a backreaction effect.  Since those are interesting
approaches, it may be useful to use their approaches to discuss the present
subject and compare those results with ours in future.

\acknowledgements
We would like to thank T. Buchert, M. Morikawa, M. Morita,
and Y. Sota for useful discussions and comments.
This work was supported partially  by the Waseda University Grant for
Special Research Projects.

\appendix

\section{backreaction in Lagrangian approximation}
Using the  Lagrangian approximation, we
estimate a backreaction term $Q_{\cal D}$.
Using the Lagrangian perturbation ${\bf S}$, we shall rewrite the
backreaction term. For convenience, first we define the functional
determinant of three functions $A({\bf q}), B({\bf q}),
C({\bf q})$:
\begin{equation}
{\cal I} (A, B, C) \equiv \frac{\partial (A, B, C)}{\partial
(q_1, q_2, q_3)} = \varepsilon_{ijk} A_{|i} B_{|j} C_{|k} \;.
\end{equation}
The invariants of the velocity gradient
$u_{i,j}$ are given in terms of functional determinants as
\begin{eqnarray}
u_{i,j} &=& \frac{1}{2J_r} \varepsilon_{jkl} {\cal J}
(\dot{r}_i, r_k, r_l) \;,\label{eq:a2} \\
{\bf I}(u_{i,j}) &=& \frac{1}{2 J_r}\varepsilon_{ijk} {\cal
J}(\dot{r}_i, r_j, r_k) \;,\label{eq:a3}  \\
{\bf II}(u_{i,j}) &=& \frac{1}{2 J_r}\varepsilon_{ijk} {\cal
J}(\dot{r}_i, \dot{r}_j, r_k) \;.
\label{eq:a4}
\end{eqnarray}
Introducing
The r.h.s. of the Lagrangian
coordinates $q_i$ and perturbations $S_i$ by $r_i=a(t)(q_i+S_i)$,
the r.h.s. of these equations are rewritten  as
\begin{eqnarray}
{\cal J}(\dot{r}_i, r_j, r_k) &=& a^2 \dot{a} \left[{\cal
J}(q_i,q_j,q_k)+3{\cal J}(S_i, q_j, q_k)
+3{\cal J}(S_i, S_j, S_k)+{\cal J}(S_i, S_j, S_k)\right]
\nonumber \\
    && +a^3\left[{\cal J}(\dot{S}_i, q_j, q_k)+2{\cal J}
    (\dot{S}_i, S_j, q_k)
+{\cal J}(\dot{S}_i, S_j, S_k)\right] \;,
\label{eqn:func_det1}\\
{\cal J}(\dot{r}_i, \dot{r}_j, r_k) &=& a\dot{a}^2\left[{\cal
J}(q_i, q_j, q_k)+3
{\cal J}(S_i, q_j, q_k)+3{\cal J}(S_i, S_j, q_k)+{\cal
J}(S_i, S_j, S_k)\right] \nonumber \\
&& +2a^2\dot a\left[{\cal J}(\dot{S}_i, q_j, q_k)+2{\cal
J}(\dot{S}_i, S_j, q_k)
+{\cal J}(\dot{S}_i, S_j, S_k)\right] \nonumber \\
&& +a^3\left[{\cal J}(\dot{S}_i, \dot{S}_j, q_k)+{\cal J}
(\dot{S}_i, \dot{S}_j, S_k) \right] \;. \label{eqn:func_det2}
\end{eqnarray}
Introducing a simplified description
follows:
\begin{eqnarray}
{\bf I}_s\equiv {\bf I}(S_{i|j})\;, \; {\bf II}_s\equiv {\bf
II}(S_{i|j})
\;, \; {\bf III}_s\equiv{\bf III}(S_{i|j}) \;, \nonumber \\
{\bf I}_{\dot s}\equiv {\bf I}(\dot S_{i|j})\;, \; {\bf
II}_{\dot s}\equiv
{\bf II}(\dot S_{i|j})\,, \; {\bf III}_{\dot s}\equiv {\bf
III}(\dot S_{i|j}) \;,
\end{eqnarray}
we find that most terms in
(\ref{eq:a2})-(\ref{eq:a4}) with
(\ref{eqn:func_det1}), (\ref{eqn:func_det2}) are given by the following
quantities;
\begin{eqnarray}
\varepsilon_{ijk} {\cal J}(q_i, q_j, q_k) = 6 \;, ~~~~&&~~~~
\varepsilon_{ijk} {\cal J}(S_i, q_j, q_k) = 2{\bf I}_s \;,
\nonumber \\
\varepsilon_{ijk} {\cal J}(S_i, S_j, q_k) = 2{\bf II}_s
\nonumber \;, ~~~~&&~~~~
\varepsilon_{ijk}  {\cal J}(S_i, S_j, S_k) = 6{\bf III}_s\nonumber
\;,\nonumber  \\
\varepsilon_{ijk} {\cal J}(\dot{S}_i, q_j, q_k) = 2{\bf
I}_{\dot s} \;,~~~~&&~~~~
\varepsilon_{ijk}
{\cal J}(\dot{S}_i, \dot{S}_j, q_k) = 2 {\bf II}_{\dot s}
\;.
\end{eqnarray}
   ${\cal J}(\dot{S}_i,  S_j, q_k)$ and ${\cal
J} (\dot{S}_i, S_j, S_k)$ are written as
\begin{eqnarray}
\varepsilon_{ijk} {\cal J} (\dot{S}_i, S_j, q_k) &=& \frac
{\partial}{\partial t} {\bf II}_s \;, \\
\varepsilon_{ijk} {\cal J} (\dot{S}_i, S_j, S_k) &=& 2\frac
{\partial}{\partial t} {\bf III}_s \;,
\end{eqnarray}
but the last term ${\cal J} (\dot{S}_i, \dot{S}_j, S_k)$ cannot be written
by any of ${\bf I}_s$, ${\bf II}_s$, ${\bf III}_s$ and those time
derivatives.

We then finally obtain
\begin{eqnarray}
J_r{\bf I}(u_{i,j}) &=& \frac{1}{2} \left\{6a^2 \dot{a}
\left(1+{\bf I}_s+{\bf II}_s+{\bf III}_s
\right)+2a^3 \left({\bf I}_{\dot{s}}+2\frac{\partial}
{\partial t}
{\bf II}_s+\frac{\partial}{\partial t}{\bf III}_s\right)
\right\} \;, \\
J_r{\bf II}(u_{i,j})  &=& \frac{1}{2}\left[6a \dot{a}^2
\left(1+{\bf I}_s+{\bf II}_s+6{\bf III}_s
\right) \right.\nonumber \\
&&\quad +4a^2\dot{a}\left(1+{\bf I}_{\dot{s}}+\frac{
\partial}{\partial t}
{\bf II}_s+\frac{\partial}{\partial t}{\bf III}_s\right)
\nonumber\\
&&\quad \left.+2a^3\left\{{\bf II}_{\dot s}+\varepsilon_{ijk}
{\cal J}(\dot{S}_i, \dot{S}_j ,S_k) \right\}\right] \;.
\nonumber \\
\end{eqnarray}

Using these invariants, we find a backreaction term in terms of the
Lagrangian perturbations as
\begin{eqnarray}
Q_{\cal D} &=& \frac{2}{3}\frac{1}{(1+\left < {\bf I}_s
\right >_{{\cal D}_i}+
\left < {\bf II}_s \right >_{{\cal D}_i}+\left < {\bf III}_s
\right >_{{\cal D}_i})^2}
\nonumber \\
&& \times \left[3(1+ \left < {\bf I}_s \right >_{{\cal D}_i}
+\left < {\bf II}_s \right >_{{\cal D}_i}
+\left < {\bf III}_s \right >_{{\cal D}_i}) \right.\nonumber
\\
&& \left(\left < {\bf II}_{\dot s} \right >_{{\cal D}_i}
+\left < \varepsilon_{ijk} {\cal J}(\dot{S}_i, \dot{S}_j
,S_k)
\right >_{{\cal D}_i} ) \right.
   \nonumber \\
&&\left.-\left( \left < {\bf I}_{\dot s} \right >_{{\cal
D}_i}
+2 \left < \frac{\partial}{\partial t}
{\bf II}_s \right >_{{\cal D}_i}+ \left < \frac{\partial}
{\partial t}
{\bf III}_s \right >_{{\cal D}_i} \right)^2\right] \;.
\label{eqn:LA_QD}
\end{eqnarray}

If we take into account the perturbations up to the second order,
we find
\begin{eqnarray}
Q_{\cal D} &=& 2 \left < {\bf II}_{\dot s} \right
>_{{\cal D}_i} - \frac{2}{3} \left < {\bf I}_{\nabla_q
\dot{\varphi}} \right
>_{{\cal D}_i}^2  ,
\label{eqn:LA_QD_2nd}
\end{eqnarray}
where we use the variables in Eqs. (\ref{psi_def}) and (\ref{zeta_def}).
Note that the second order perturbation $\zeta $ do not appear
in Eq. (\ref{eqn:LA_QD_2nd}).

Buchert et al estimated the backreaction
term $Q_{\cal D}$ assuming  ZA, i.e. $a_b=a_H$ and the  Lagrangian
perturbation ${\bf S}$ is separable as
\begin{equation}
{\bf x} = {\bf q} + b(t) \nabla_q \varphi ({\bf q}) \;,
\end{equation}
and found the similar form to (\ref{eqn:LA_QD}).\cite{BucKerSic00}

\end{document}